# BEAM DYNAMICS IN A DOUBLE-CHANNEL WAVEGUIDE WITH GAIN AND LOSS

PENGFEI LI, RUJIANG LI, LU LI*

Institute of Theoretical Physics, Shanxi University, Taiyuan 030006, China

*E-mail: llz@sxu.edu.cn

*Abstract*. A double-channel waveguide with gain and loss is addressed and the corresponding coupled-mode equations are established by employing the coupled mode approach. Based on the coupled-mode equations, the beam dynamics in the double-channel waveguide with gain and loss is investigated, and the results show that there exist three distinct dynamical behaviors, which are amplification, attenuation to zero and tending to a steady value (or equilibrium state), respectively. Finally, it is shown that the theoretical results suggested by the coupled-mode equations agree well with the numerical simulations.

## 1. INTRODUCTION

Since the concept of parity-time (PT) symmetry was applied to construct complex extension of quantum mechanics [1-3], physical systems exhibiting PT symmetry have attracted much more attention and have been extensively studied in recent years [4-6]. Particularly, in optics, PT-related notions can be implemented in PT- symmetric coupler [7-12] and PT-symmetric optical lattices [13-18], and the experimental observations have been demonstrated [19-21]. Thus, optics can provide a fertile ground to investigate PT-related beam dynamics including the non-reciprocal responses, the power oscillations, and the optical transparency. Recently, based on the properties in the PT symmetry breaking region, different optical components have been theoretically and experimentally realized [22-26]. As a natural extension, in this paper, we investigate the optical beam dynamics in a double-channel waveguide with gain and loss, and suggest three distinct dynamical behaviors, which are amplification, attenuation to zero and tending to a steady value (or equilibrium state), respectively. These results are verified by numerical simulations.

## 2. MODEL AND REDUCTIONS

In the context of the paraxial theory of diffraction by involving index guiding and a gain/loss profile, the electric field envelope obeys a normalized Schrödinger equation as follows

$$i\frac{\partial \phi}{\partial z}+\frac{1}{2k}\frac{\partial^2 \phi}{\partial x^2}+U(x)\phi=0. \tag{1}$$

Here $\phi=\phi(x,z)$ is the complex envelope of the electric field, $z=Z/(2kx_0^2)$ is a scaled propagation distance, and $x=X/x_0$ is a dimensionless transverse coordinate, where $x_0$ is an arbitrary spatial scale and $k=2\pi n_0/\lambda_0$ is the wave number with $n_0$ being the background refractive index and $\lambda_0$ being the wavelength of the optical source generating the beam. The function $U(x)=2k^2x_0^2[n(x)-n_0]/n_0 \equiv V(x)+iW(x)$ represents the normalized complex index distribution, in which the refractive index profile $V(x)$ and the gain/loss profile $W(x)$ are of the form $V(x)=V_1(x)+V_2(x)$ and $W(x)=W_1(x)+W_2(x)$, where $V_1(x)$ and $V_2(x)$ equal to $V_0$ in the regions of $-d-D/2<x<-D/2$ and $D/2<x<d+D/2$, respectively, and $V_1(x)=V_2(x)=0$ otherwise, and $W_1(x)=-W_1$ and $W_2(x)=W_2$ in the regions of $-d-D/2<x<-D/2$ and $D/2<x<d+D/2$, respectively, otherwise they are zero, as shown in Fig. 1. Here $V_0$ is the modulation depth of the refractive index, and $W_1$ and $W_2$ are the dimensionless gain or loss parameters. Without loss of generality we assume that $W_1>0$ and $W_2>0$, which implies that the left-channel in Fig. 1 is a gain-guiding waveguide while the right-channel is a loss-guiding waveguide, forming a double-channel waveguide with gain and loss. Especially, when $W_1=W_2$, the complex index distribution $U(x)$ satisfies the PT symmetric condition and has been extensively studied [7-12]. Here, we will discuss the case $W_1 \neq W_2$, and explore the corresponding optical behavior and characteristics.

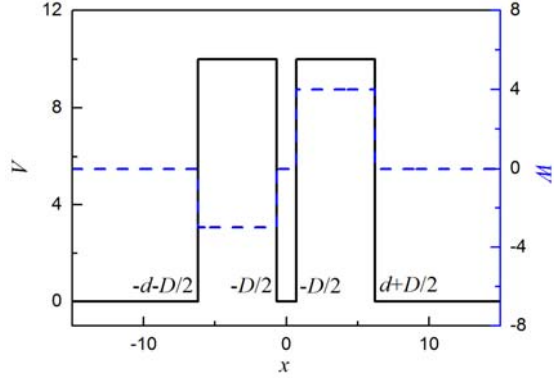

Fig. 1- The real part (black solid) and the imaginary part (blue dashed) of the refractive index profile of a double-channel waveguide structure with gain and loss. Here $D$ is the separation between the double-channel waveguides, and $d$ represents the thickness of the waveguide core.

In order to investigate the dynamical characteristic of the beam propagation in such waveguide, we will employ the coupled-mode approach. We assume that the solution of Eq. (1) can be expressed as a superposition of the local modes of the individual channels without gain and loss as follows [7]

$$\phi(x,z)=\left[a(z)u_1(x)+b(z)u_2(x)\right]e^{i\beta z}, \qquad (2)$$

where $a(z)$ and $b(z)$ represent the field amplitudes in left- and right-channel waveguides, respectively, and $u_j=u_j(x)$ satisfies $d^2u_j/dx^2+V_j(x)u_j=\beta u_j$, $j=1,2$. By substituting Eq. (2) into Eq. (1) and multiplying by $u_2(-x)$ and $u_1(-x)$ respectively, and by integrating over the whole space for the variable $x$, we obtain a set of equations for the field amplitudes $a(z)$ and $b(z)$ in the form

$$i\frac{da(z)}{dz}+(\delta-i\gamma_a)a(z)+(\kappa-i\sigma_a)b(z)=0, \qquad (3)$$

$$i\frac{db(z)}{dz}+(\delta+i\gamma_b)b(z)+(\kappa+i\sigma_b)a(z)=0. \qquad (4)$$

Here $\delta=(I_{12}J_{121}-I_{11}J_{122})/\Delta$ represents the phase shift and $\kappa=(I_{12}J_{122}-I_{11}J_{121})/\Delta$ is the real part of the scaled coupling coefficient, which are determined by the refractive index; $\gamma_a=[I_{11}(C_{111}+C_{211})-I_{12}(C_{212}+C_{112})]$ and $\gamma_b=[I_{12}(C_{121}+C_{221})-I_{11}(C_{122}+C_{222})]$ are the effective gain/loss coefficients, and $\sigma_a=[I_{11}(C_{121}+C_{221})-I_{12}(C_{122}+C_{222})]/\Delta$ and $\sigma_b = [I_{12}(C_{111} + C_{211}) - I_{11}(C_{212} + C_{112})]/\Delta$ are the imaginary parts of the scaled coupling coefficients, respectively, which are determined by the gain/loss distribution, where $\Delta = I_{12}^2 - I_{11}^2$,

$I_{mj} = \int_{-\infty}^{+\infty} u_m(x)u_j(-x)dx$, $J_{kmj} = \int_{-\infty}^{+\infty} V_k(x)u_m(x)u_j(-x)dx$, and $C_{kmj} = \int_{-\infty}^{+\infty} W_k(x)u_m(x)u_j(-x)dx$,

$k, m, j = 1, 2$. Note that we have used the relations $I_{22} = I_{11}$, $I_{21} = I_{12}$, $J_{212} = J_{121}$, and $J_{211} = J_{122}$. Thus we presented a set of coupled-mode equations which can be used to describe the optical beam dynamics in the double-channel waveguide with gain and loss. Comparing with the corresponding theory of coupled optical PT-symmetric structures reported in Ref. [7] (the special case of $W_1 = W_2$), the local eigenfunction $u_j(x)$ in Eq. (2) satisfies the eigenvalue problem for a Hermitian Hamiltonian, i. e., $d^2 u_j / dx^2 + V_j(x)u_j = \beta u_j$, $j = 1, 2$, which does not include the imaginary part $W_j(x)$ of the complex index distribution $U(x)$ and so is more easy to treat. Thus, we can investigate the optical beam dynamics in the double-channel waveguide with gain and loss by employing the coupled-mode equations.

### 3. DYNAMIC CHARACTERISTICS

In this section, we will analyze the optical beam dynamics in the double-channel waveguide with gain and loss. Firstly we consider the case of $W_1 < W_2$, which implies that $\sigma_a > \sigma_b$ and $\gamma_a < \gamma_b$. Thus one can obtain the solution for Eqs. (3) and (4) in the form

$$\begin{pmatrix} a(z) \\ b(z) \end{pmatrix} = e^{(i\delta - \gamma_-)z} \left[ \mathbf{A} e^{(\chi + i\omega)z} + \mathbf{B} e^{-(\chi + i\omega)z} \right] \begin{pmatrix} a_0 \\ b_0 \end{pmatrix}, \quad (5)$$

where $a_0$ and $b_0$ represent the initial state, and **A** and **B** are 2×2 matrices, which are given by

$$\mathbf{A} = \begin{pmatrix} \dfrac{(\chi + i\omega) + \gamma_+}{2(\chi + i\omega)} & \dfrac{i(\kappa - i\sigma_a)}{2(\chi + i\omega)} \\ \dfrac{i(\kappa + i\sigma_b)}{2(\chi + i\omega)} & \dfrac{(\chi + i\omega) - \gamma_+}{2(\chi + i\omega)} \end{pmatrix}$$

$$\mathbf{B} = \begin{pmatrix} \dfrac{(\chi + i\omega) - \gamma_+}{2(\chi + i\omega)} & -\dfrac{i(\kappa - i\sigma_a)}{2(\chi + i\omega)} \\ -\dfrac{i(\kappa + i\sigma_b)}{2(\chi + i\omega)} & \dfrac{(\chi + i\omega) + \gamma_+}{2(\chi + i\omega)} \end{pmatrix}$$

and $\gamma_\pm = (\gamma_b \pm \gamma_a)/2$, $\omega = \left\{ \left[ (\mu^2 + \nu^2)^{1/2} + \mu \right]/2 \right\}^{1/2}$, and $\chi = \left\{ \left[ (\mu^2 + \nu^2)^{1/2} - \mu \right]/2 \right\}^{1/2}$ with $\mu = \kappa^2 + \sigma_a \sigma_b - \gamma_+^2$ and $\nu = \kappa(\sigma_b - \sigma_a)$. Similarly, for the case of $W_1 > W_2$, which implies that $\sigma_a < \sigma_b$ and $\gamma_a > \gamma_b$, the result is the same except $\omega$ is replaced by $-\omega$. It should be pointed out that when $W_1 = W_2$, i. e., $\sigma_a = \sigma_b$ and $\gamma_a = \gamma_b$, the above results hold too. Indeed, in this case, the model (1) can describe the beam propagation in the PT-symmetric double-channel waveguide [10]. It has been shown that there exists a critical point $\gamma_a = \sqrt{\kappa^2 + \sigma_a^2}$ such that below this critical point, i. e., $\gamma_a < \sqrt{\kappa^2 + \sigma_a^2}$, the eigenvalues for the eigenvalue problem of the coupled-mode equations (3) and (4) are the real numbers $\lambda_\pm = \delta \pm \omega$ and so the optical beam dynamics exhibits an oscillatory evolution behavior because $\chi = 0$. While as $\gamma_a > \sqrt{\kappa^2 + \sigma_a^2}$ the eigenvalues for the eigenvalue problem of Eqs. (3) and (4) become the complex numbers $\lambda_\pm = \delta \pm i\chi$, which means that a phase transition occurs, hence the corresponding total power is exponentially increasing.

In the following, we discuss the case of $W_1 \neq W_2$. In this case, the eigenvalues for the eigenvalue problem of Eqs. (3) and (4) are always complex-valued. Hence the beam dynamics may exhibit some different characteristics. When $W_1 < W_2$, Eq. (5) can present the total power as follows

$$|a(z)|^2 + |b(z)|^2 = e^{-2\gamma_- z} \begin{pmatrix} a_0^* & b_0^* \end{pmatrix} \left[ \mathbf{A}^\dagger \mathbf{A} e^{2\chi z} + \mathbf{B}^\dagger \mathbf{A} e^{i2\omega z} + \mathbf{A}^\dagger \mathbf{B} e^{-i2\omega z} + \mathbf{B}^\dagger \mathbf{B} e^{-2\chi z} \right] \begin{pmatrix} a_0 \\ b_0 \end{pmatrix}, \qquad (6)$$

where '†' represents the transpose conjugation and '∗' represents the complex conjugation. From Eq. (6) it can be seen that the parameter $\omega$ represents the oscillatory frequency, while the parameters $\gamma_-$ and $\chi$ are the amplification/attenuation factors, which can be used to characterize the asymptotic behavior of the optical beam. Also one find that the final three terms in the right-hand side of Eq. (6) tend to zero as $z \to \infty$ due to $\gamma_- > 0$, therefore the total power is exponentially increasing (decreasing) for $\gamma_- < \chi$ ($\gamma_- > \chi$), while when $\gamma_- = \chi$, the total power tends eventually to a steady value $(a_0^*, b_0^*) \mathbf{A}^\dagger \mathbf{A} (a_0, b_0)^T$ as $z \to \infty$, forming an equilibrium state, where the superscript '$T$' represents the transpose. However, for the case of $W_1 > W_2$ [note that ω in Eq. (6) should be replaced by $-\omega$ as $W_1 > W_2$], there exists no such a steady value and the total power is exponentially amplified due to $\gamma_- < 0$. Thus it can be suggested that there are three distinct evolution characteristics in the double-channel waveguide with gain and loss, which are amplification, attenuation to zero and approaching a steady value (or equilibrium state), respectively.

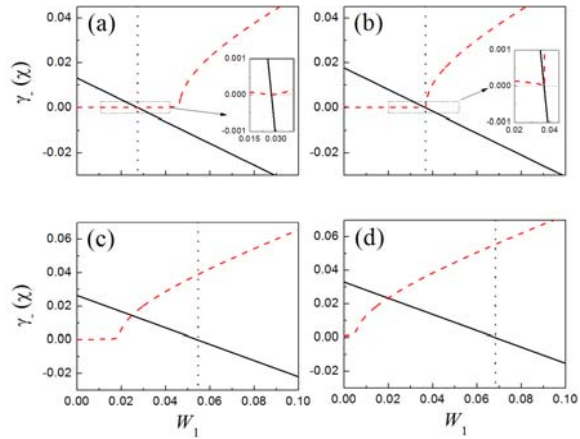

Fig. 2- The dependence of $\gamma_-$ (the black solid curves) and $\chi$ (the red dashed curves) on $W_1$ for (a) $W_2$=0.02733; (b) $W_2$=0.0367; (c) $W_2$=0.05465 and (d) $W_2$=0.06832, where the vertical dotted lines represent $W_1$=$W_2$. Here the system parameters are $d$=4μm, $D$=4 μm, and $V_0$=2.5.

In order to determine the parameter range for the three distinct beam dynamics, we present the dependence of $\gamma_-$ and $\chi$ on $W_1$ for the different $W_2$, as shown in Fig. 2, where the parameters are taken as $\lambda_0 = 0.5145\ \mu m$, $n_0 = 2.797$, $x_0 = 1\ \mu m$, $W_2 = 0.02733$, $0.0367$, $0.05465$, and $0.06832$, which correspond to the actual loss parameters as $4\ cm^{-1}$, $5.372\ cm^{-1}$, $8\ cm^{-1}$, and $10\ cm^{-1}$, respectively.

From Figs. 2(a) and 2(b), one can see that when $W_2 \leq 0.0367$ the crosspoint of $\gamma_-$ and $\chi$ equals to zero, which corresponds to $W_1 = W_2$ [see the inset in Figs. 2(a) and 2(b)], and we have $\gamma_- > \chi$ in the region of $W_1 < W_2$ and $\gamma_- < \chi$ in the region of $W_1 > W_2$. However, when $W_2 > 0.0367$, the crosspoint of $\gamma_-$ and $\chi$ is located in the region of $W_1 < W_2$, as shown in Figs. 2(c) and 2(d). This fact means that there exists a critical value $W_2^{crit} = 0.0367$ (for our choice of the parameters) so that when $W_2 > W_2^{crit}$, the cases of $\gamma_- > \chi$, $\gamma_- = \chi$, and $\gamma_- < \chi$ can appear simultaneously for $W_1 < W_2$.

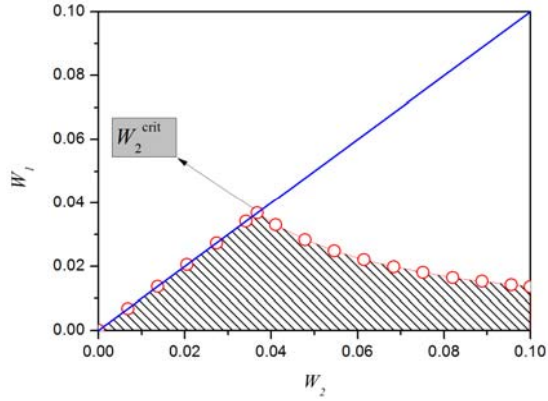

Fig. 3- The range of the three distinct evolution behaviors in $W_2W_1$-plane. The blue line represents $W_1=W_2$. The red circles represent $\gamma_- = \chi$. Here the system parameters are the same as in Fig. 2.

Figure 3 summarizes the parameter range for the three distinct beam dynamics in $W_2W_1$-plane. From Fig. 3 one can see that there exists a curve with $\gamma_- = \chi$ (see the red circle curve in Fig. 3), on which the optical beam evolution behavior is oscillatory or tends eventually to a steady value. Indeed, the curve is consisted of two parts, one is located on the line of $W_1 = W_2$ on which the beam evolution behavior is oscillatory, the other is located in the region of $W_1 < W_2$ on which the beam dynamics eventually tends to a steady value.

Therefore, the critical value $W_2^{crit}$ should be the PT-symmetry-breaking point. Furthermore, one can see that the curve present a boundary, above which the beam dynamics exhibits a exponential amplification in the oscillatory form (the white region) and below which the beam dynamics is attenuated in the oscillatory form (the shadow region). Thus one can implement the distinct evolution behaviors of the optical beam in the double-channel waveguide with gain and loss by properly choosing the parameters $W_1$ and $W_2$.

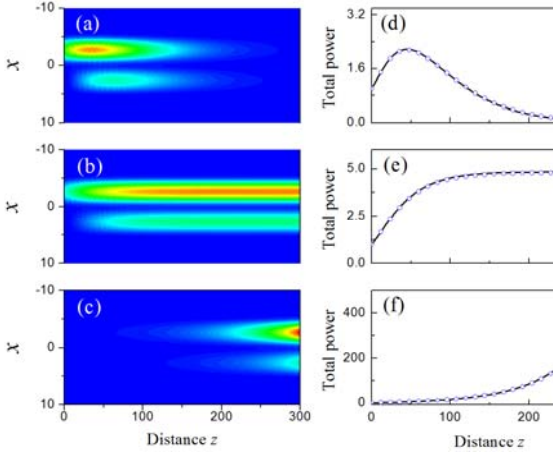 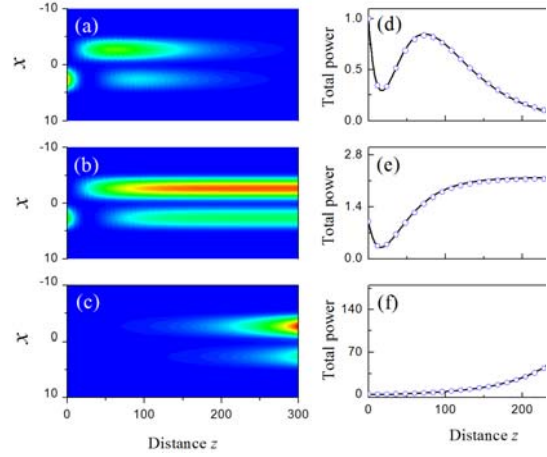

Fig. 4- (a), (b) and (c) are the evolution plots of the optical intensity given by Eq. (1) with the initial state $u_1(x)$ for $W_1 = 0.02$, $0.024765$, and $0.03$ respectively. (d), (e) and (f) are the corresponding total power, where the black curves are the analytical results given by Eq. (6) and the blue circles are the numerical results by simulating Eq. (1) with the initial state $u_1(x)$. Here $W_2 = 0.05465$ and the other parameters are the same as in Fig. 2.

Fig. 5- (a), (b) and (c) are the evolution plots of the optical intensity given by Eq. (1) with the initial state $u_2(x)$ for $W_1 = 0.02$, $0.024765$, and $0.03$ respectively. (d), (e) and (f) are the corresponding total power, where the black curves are the analytical results given by Eq. (6) and the blue circles are the numerical results by simulating Eq. (1) with the initial state $u_2(x)$. Here $W_2 = 0.05465$ and the other parameters are the same as in Fig. 2.

In the following, we will verify the three distinct evolution dynamics by directly simulating Eq. (1) with the initial states $u_1(x)$ and $u_2(x)$, which correspond to the initial excitation state (1,0) and (0,1) for Eqs. (3) and (4), respectively.

The results are summarized in Fig. 4 and Fig. 5, respectively, where $W_2 = 0.05465$, corresponding to the loss parameter $8\ cm^{-1}$, $W_1 = 0.02$, $0.024765$, and $0.03$, and to the cases of $\gamma_- > \chi$, $\gamma_- = \chi$, and $\gamma_- < \chi$, respectively.

Three distinct asymptotic behaviors are exhibited: Figs. 4(a) and Fig. 5(a) indicate the oscillatory decaying behavior, Figs. 4(c) and Fig. 5(c) present the exponential amplification behavior, while Figs. 4(b) and Fig. 5(b) show that the optical beam tends to a equilibrium state with an increasing of the propagation distance, which corresponds to the case of $\gamma_- = \chi$. Also, Figs. 4(d)-(f) and Figs. 5(d)-(f) present the evolution plots of the total power, and the results show that the analytical values agree well with the corresponding numerical ones.

Furthermore, we also performed calculations for the equilibrium state for different values of $W_2$. The obtained results show that the steady value is decreasing with the increase of $W_2$, as illustrated in Fig. 6, and the theoretical analytical results agree well with the corresponding numerical ones.

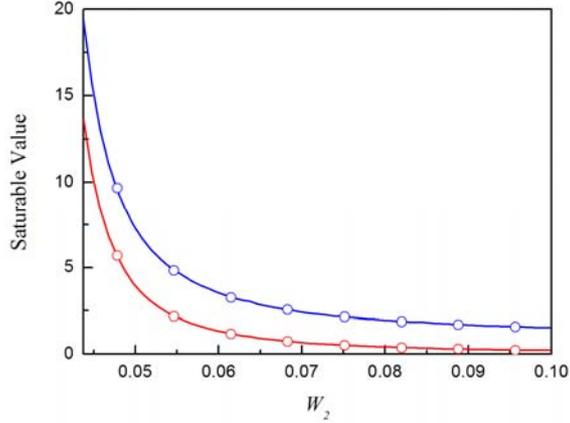

Fig. 6 The dependence of the steady value on $W_2$. Blue and red curves represent the analytical results corresponding to the initial excitation states (1, 0) and (0, 1), respectively. Blue circles and red circles are the numerical results. Here the system parameters are the same as in Fig. 2.

## 4. CONCLUSIONS

In summary, we have presented the coupled-mode equations for a double-channel waveguide with gain and loss by employing the coupled mode theory. Based on the coupled-mode equations, we have suggested that there exist three distinct dynamical behaviors, which are exponential amplification, exponential attenuation to zero and tending to a steady value, respectively, and we have presented their parameter range. Finally, it has been shown that the theoretical results suggested by the coupled-mode equations agree well with the corresponding numerical simulations. These unique properties may be used to control the transmission behavior of the optical beam by properly choosing system's parameters.

## 5. ACKNOWLEDGEMENT

This research is supported by the National Natural Science Foundation of China grant 61078079 and the Shanxi Scholarship Council of China Grant No. 2011-010.

## 6. REFERENCES

[1] C. M. Bender, S. Boettcher, *Real spectra in non-Hermitian Hamiltonians having PT-symmetry,* Phys. Rev. Lett., **80**, 5243-5246 (1998).

[2] C. M. Bender, S. Boetccer, P. N. Meisinger, *PT-symmetric quantum mechanics*, J. Math. Phys., **40**, 2201-2229 (1999).

[3] C. M. Bender, D. C. Brody, H. F. Jones, *Complex extension of Quantum Mechanics*, Phys. Rev. Lett., **89**, 270401 (2002).

[4] C. M. Bender, *Making sense of non-Hermitian Hamiltonians*, Rep. Prog. Phys., **70**, 947 (2007).

[5] A. Mostafazadeh, *Quantum brachistochrone problem and the geometry of the state space in pseudo-Hermitian quantum mechanics,* Phys. Rev. Lett., **99**, 130502 (2007).

[6] A. Mostafazadeh, *Spectral singularities of complex scattering potentials and reflection and transmission coefficients at real energies*, Phys. Rev. Lett., **102**, 220402 (2009).

[7] R. El-Ganainy, K.G. Makris, D. N. Christodoulides, Z. H. Musslimani, *Theory of coupled optical PT-symmetric structures*, Opt. Lett., **32**, 2632-2634 (2007).

[8] S. Klaiman, U. Günther, N. Moiseyev, *Visualization of branch point in PT-symmetric waveguides*, Phys. Rev. Lett., **101**, 080402 (2008).

[9] T. Kottos, *Broken symmetry makes light work*, Nat. Phys., **6**, 166-167 (2010).

[10] Li Chen, Rujiang Li, Na Yang, Da Chen, Lu Li, *Optical modes in PT-symmetric double-channel waveguides*, Proc. Romanian Acad. A, **13**, 46 (2012).

[11] R. Driben, B. A. Malomed, *Stabilization of solitons in PT models with supersymmetry by periodic management*, Europhys. Lett. (EPL), **96**, 51001 (2011).

[12] R. Driben, B. A. Malomed, *Stability of solitons in parity-time-symmetric couplers*, Opt. Lett., **36**, 4323-4325 (2011).

[13] Z. H. Musslimani, K. G. Makris, R. El-Ganainy, D. N. Christodoulides, *Optical Solitons in PT-periodic potentials*, Phys. Rev. Lett., **100**, 030402 (2008).

[14] K. G. MAKRIS, R. El-GANAINY, D. N. Christodoulides, Z. H. Musslimani, *Beam dynamics in PT-symmetric optical lattices*, Phys. Rev. Lett., **100**, 103904 (2008).

[15] K. G. Makris, R. El-Ganainy, D. N. Christodoulides, Z. H. Musslimani, *PT -symmetric optical lattices*, Phys. Rev. A, **81**, 063807 (2010).

[16] Chunyan Li, Haidong Liu, Liangwei Dong, *Multi-stable solitons in PT-symmetric optical lattices*, Opt. Express., **20**, 16823-16831 (2012).

[17] Yingji He, Xing Zhu, D. Mihalache, Jinglin Liu, Zhanxu Chen, *Solitons in PT–symmetric optical lattices with spatially periodic modulation of nonlinearity*, Opt. Commun., **285**, 3320-2224 (2012); Y. He, X. Zhu, D. Mihalache, J. Liu, Z. Chen, *Lattice solitons in PT-symmetric mixed linear-nonlinear optical lattices*, Phys. Rev. A, **85**, 013831 (2012); Y. He, D. Mihalache, X. Zhu, L. Guo, Y. V. Kartashov, *Stable surface solitons in truncated complex potentials*, Opt. Lett., **37**, 2526-2528 (2012); Yingji He, D. Mihalache, *Spatial solitons in parity-time-symmetric mixed linear-nonlinear optical lattices: recent theoretical results*, Rom. Rep. Phys., **64**, xxx (2012).

[18] Chengping Yin, Yingji He, Huagang Li, Jianing Xie, *Solitons in parity-time symmetric potentials with spatially modulated nonlocal nonlinearity*, Opt. Express, **20**, 19355-19362 (2012).

[19] A. Guo, G. J. Salamo, D. Duchesne, R. Morandotti, M. Volatier-Ravat, V. Aimez, G. A. Siviloglou, D. N. Christodoulides, *Observation of PT-symmetry breaking in complex optical potentials*, Phys. Rev. Lett., **103**, 093902 (2009).

[20] E. Rüter, K. G. Makris, R. El-Ganainy, D. N. Christodoulides, M. Segev, D. Kip, *Observation of parity-time symmetry in optics*, Nat. Phys., **6**, 192-195 (2010).

[21] A. Regensburger, C. Bersch, M.-A. Miri, G. Onishchukov, D. N. Christodoulides, U. Peschel, *Parity-time synthetic photonic lattices*, Nature, **488**, 167-171 (2012).

[22] L. Feng *et al.*, *Nonreciprocal light propagation in a silicon photonic circuit*, Science, **333**, 729-733 (2011).

[23] F. Nazari, M. Nazari, M. K. Moravvej-Farshi, *A 2 × 2 spatial optical switch based on PT-symmetry*, Opt. Lett., **36**, 4368-4370 (2011).

[24] M. A. Miri, P. Li Kam Wa, D. N. Christodoulides, *Large area single-mode parity–time-symmetric laser amplifiers*, Opt. Lett., **37**, 764-766 (2012).

[25] H. Ramezani, T. Kottos, R. El-Ganainy, D. N. Christodoulides, *Unidirectional nonlinear PT-symmetric optical structures*, Phys. Rev. A, **82**, 043803 (2010).

[26] Z. Lin, H. Ramezani, T. Eichelkraut, T. Kottos, H. Cao, D. N. Christodoulides, *Unidirectional invisibility induced by PT-symmetric periodic structures*, Phys. Rev. Lett., **106**, 213901 (2011).